\documentclass[aps,prl,twocolumn,superscriptaddress,amssymb]{revtex4}

\usepackage{color}
\usepackage{graphicx}
\usepackage{epstopdf}

\newcommand\lsim{\mathrel{\rlap{\lower4pt\hbox{\hskip1pt$\sim$}}
    \raise1pt\hbox{$<$}}}
\newcommand\gsim{\mathrel{\rlap{\lower4pt\hbox{\hskip1pt$\sim$}}
    \raise1pt\hbox{$>$}}}
\def\bea{\begin{eqnarray}}
\def\eea{\end{eqnarray}}
\def\ba{\begin{array}}
\def\ea{\end{array}}
\def\bc{\begin{center}}
\def\ec{\end{center}}
\def\nn{\nonumber}

\def\f{\frac}

\def\f#1#2{\frac{#1}{#2}}

\begin{document}

\preprint{SNUTP 05-08}

\title{Radiatively Generated Maximal Mixing Scenario for the Higgs Mass
\\ and the Least Fine Tuned Minimal Supersymmetric Standard Model}

\author{Radovan Derm\' \i\v sek}

\affiliation{School of Natural Sciences, Institute for Advanced Study, Princeton,
NJ 08540, U.S.A.}

\author{Hyung Do Kim}

\affiliation{School of Physics and Center for Theoretical Physics, 
Seoul National University, Seoul, 151-747, Korea}

\date{January 5, 2006}

\begin{abstract}

We argue that given the experimental constraints on the Higgs mass the least fine tuned 
parameter space of minimal supersymmetric standard model is with negative stop masses squared at the 
grand unification scale. 
While stop mass squared is typically driven to positive values at the weak scale,
the contribution to the Higgs mass squared parameter from the running can be arbitrarily small,  
which reduces fine tuning of electroweak symmetry breaking.
At the same time the stop mixing is necessarily  enhanced and the maximal mixing scenario for 
the Higgs mass can be generated radiatively even when starting with negligible mixing at the 
unification scale. This highly alleviates constraints on possible models for supersymmetry breaking 
in which fine tuning is absent.

\end{abstract}


\maketitle




Minimal Supersymmetric Standard Model  (MSSM) is a promising candidate
for describing physics above the electroweak (EW) scale. The three gauge
couplings unify at the GUT (grand unified theory) scale $\sim~2~\times~10^{16}$~GeV
within a few percent, and 
the hierarchy between the EW scale and the GUT scale is naturally stabilized by
supersymmetry (SUSY).
In addition, if we add soft-supersymmetry-breaking terms at the GUT scale
we typically find that the mass squared of the Higgs doublet which couples to the 
top quark ($H_u$), is driven to negative values at the EW scale. This triggers 
electroweak symmetry breaking and the EW scale is naturally understood from 
SUSY breaking scale.
Furthermore, assuming R-parity, the lightest supersymmetric
particle (LSP) is stable and it is a natural candidate for
dark matter of the universe.

The real virtue of supersymmetry is that the above mentioned features do not 
require any specific relations between soft-supersymmetry-breaking parameters (SSBs) 
and the only strong requirement on SUSY breaking scenarios is that these terms 
are of order  the EW scale. However generic SSBs near the EW scale generically 
predict too light Higgs mass which is ruled out by LEP limits. The exact value of 
the Higgs mass is not relevant for low energy physics, nothing crucially depends 
on it, and yet, in order to stay above LEP limits ($m_h \gsim 114.4$ 
GeV~\cite{Higgs_MSSM}) the SSBs 
have to be either considerably above the EW scale or related to each other in a 
non-trivial way. SSBs can no longer be just generic which leads to strong 
requirements 
on possible models for SUSY breaking should these provide a natural explanation for 
the scale where electroweak symmetry is broken. 

In this letter we show that such constraints are highly alleviated and the fine 
tuning is in principle
absent in scenarios which have negative stop masses squared  at the 
unification scale.
While stop mass squared is typically driven to positive values at the EW scale 
by gluino loops through
renormalization group (RG) running, the contribution to the Higgs mass squared 
parameter 
from the running (mostly due to top Yukawa coupling) can be arbitrarily small, which reduces fine tuning of electroweak symmetry breaking.  At the same 
time the stop mixing is necessarily  enhanced which is known to  enlarge the Higgs mass.
Even the maximal mixing scenario for 
the Higgs mass can be radiatively generated (starting with negligible mixing 
at the GUT scale). Thus in the least fine tuned scenarios the Higgs mass is highly enhanced  without any further assumptions. 

In spite of having tachyonic scalar masses at a high scale such scenarios are not 
excluded by our current knowledge of cosmology. We discuss constraints from charge 
and color breaking minima on possible scenarios. Finally, we discuss a  typical 
spectrum of these scenarios which is characterized by light stop, light higgsino 
and a fairly light gluino.


The tension between the direct search bound on the Higgs mass and naturalness 
of electroweak symmetry breaking can be summarized as follows~\cite{Chung:2003fi}. 
At tree level, the mass of the lightest  Higgs boson in MSSM is bounded from above by 
the mass of the Z boson,
\begin{equation}
m_{h}^2 < M_Z^2  \cos^2 2 \beta,
\label{eq:mh_tree}
\end{equation}
where $\tan \beta = v_u/v_d$ is the ratio of the vacuum expectation values 
of $H_u$ and $H_d$.
The dominant one loop correction, in case the stop mixing parameter is small, 
is proportional to $m_t^4 \log ( m_{\tilde{t}}^2 /m_t^2)$ (for simplicity we assume $m_{\tilde{t}} \simeq m_{\tilde{t}_L} \simeq m_{\tilde{t}_R}$ throughout this paper). It depends only logarithmically on stop masses 
and it has to be large in order to push the Higgs mass above the LEP limit.
A two loop calculation (we use {\it FeynHiggs 2.2.10}~\cite{Heinemeyer:1998yj,Heinemeyer:1998np} with 
$m_t = 172.7$ GeV) reveals the stop masses have to be $\gsim 900$ GeV.
 On the other hand, the mass of the Z boson ($M_Z \simeq 91$ GeV) 
is given from the minimization of the scalar potential as (for $\tan \beta \gsim 5$)
\bea \f{M_Z^2}{2} & \simeq & - \mu^2 (M_Z)
-m_{H_u}^2 (M_Z),
\label{eq:MZ}
\eea 
and the large stop masses directly affect the running of soft
scalar mass squared for $H_u$, 
\bea \delta m_{H_u}^2
& \simeq & -\f{3}{4\pi^2} m_{\tilde{t}}^2 \log
\f{\Lambda}{m_{\tilde{t}}}.  
\label{eq:del_mhu}
\eea
Numerically the loop factor times large log is of order one for
$\Lambda \sim M_{\rm GUT}$ and we have $\delta m_{H_u}^2 \simeq -
m_{\tilde{t}}^2$.
Starting with negligible $m_{H_u}^2$ at the GUT scale we find  $m_{H_u}^2 (M_Z) \simeq \delta m_{H_u}^2 
\simeq - m_{\tilde{t}}^2 \simeq -(900 \ {\rm GeV})^2$ and the correct Z mass requires 
that $\mu^2 (M_Z)$ is tuned to $m_{H_u}^2 (M_Z)$ with better than one percent accuracy.
Alternatively, we can start from large positive
$m_{H_u}^2 (M_{\rm GUT}) \sim (900 \ \rm{GeV})^2$ in which case 
$m_{H_u}^2 (M_Z) \sim - M_Z^2$ is possible. However, in this case 
the fine tuning is hidden in
$m_{H_u}^2 (M_{\rm GUT})$. Small change of the boundary condition
$m_{H_u}^2 (M_{\rm GUT})$ would generate very different value for the
EW scale and the situation is quite similar to the tuning of $\mu$. 

The situation highly improves when considering large mixing in the stop sector. The mixing is controlled by  
the ratio of $A_t - \mu \cot \beta$ and $m_{\tilde{t}}$. Since we consider parameter space where $\mu$ is 
small to avoid fine tuning and $\tan \beta \gsim 5$ in order to maximize the tree level 
Higgs mass~(\ref{eq:mh_tree}), the mixing is simply given by $A_t / m_{\tilde{t}}$.
It was realized that mixing $A_t(M_Z) / m_{\tilde{t}} (M_Z) \simeq \pm 2$ maximizes the Higgs mass for 
given $m_{\tilde{t}}$~\cite{Carena:2000dp}, while still satisfying constraints to avoid charge and color breaking (CCB) minima~\cite{LeMouel:2001ym}. 
Using {\it FeynHiggs 2.2.10} we find that $m_{\tilde{t}} (M_Z) \simeq 300$ GeV and $|A_t (M_Z)| = 450$ 
GeV (for $\tan \beta \gsim 50$),  $|A_t (M_Z)| = 500$ (for any $\tan \beta \gsim 8$) or  $|A_t (M_Z)| = 600$ GeV (for  $\tan \beta$ as small as 6) satisfies the LEP limit on the Higgs 
mass~\footnote{Slightly smaller $m_{\tilde{t}}$ is allowed, 
the actual minimal possible value of  $m_{\tilde{t}}$  
is not important for our discussion.}.  Therefore large stop mixing, $|A_t(M_Z) / m_{\tilde{t}} (M_Z)| \gsim 1.5$ is crucial for satisfying the LEP limit  with light stop masses (the physical stop mass in this case can be as small as current experimental bound, $m_{\tilde{t}_1} \gsim 100$ GeV).
 Decreasing the mixing requires increasing of $m_{\tilde{t}}$ and finally we end up with 
 $m_{\tilde{t}} \gsim 900$ GeV for small mixing.
 
In order to discuss fine tuning in this case, the approximate solution of RG equation for $m_{H_u}^2$, 
Eq.~(\ref{eq:del_mhu}), is not sufficient.
For given $\tan \beta$ we can solve RG equations exactly and express EW values of $m_{H_u}^2$, 
$\mu^2$, and consequently  
$M_Z^2$ given by Eq. (\ref{eq:MZ}), as functions of all GUT scale parameters~\cite{Ibanez:1983di,Carena:1996km}.
For $\tan \beta =10$, we have:
\bea M_Z^2 & \simeq & -1.9 \mu^2 + 5.9 M_3^2 -1.2 m_{H_u}^2 + 1.5 m_{\tilde{t}}^2 \nn \\
&&  - 0.8 A_t M_3 + 0.2 A_t^2 , 
\label{eq:MZ_gut}
\eea 
where 
parameters 
appearing on the right-hand side are the GUT scale parameters, we do not write the scale 
explicitely.
The contribution of $M_2$ to the 
above formula 
is small and when $M_2 \sim M_3$ it cancels between $\simeq - 0.4 M_2^2$ term and the mixed 
$\simeq 0.4 M_3 M_2$ term. Other scalar masses and $M_1$ appear with negligible coefficients and 
we neglect them in our discussion.
The coefficients in this expression depend only on 
$\tan \beta$ (they do not change dramatically when varying $\tan \beta$ between 5 and 50) and 
$\log (M_{GUT}/M_Z)$. 

Let us also express the EW scale values of stop mass squared, gluino mass and top trilinear 
coupling for $\tan \beta = 10$ in a similar way:
\bea
m_{\tilde{t}}^2 (M_Z) & \simeq & 5.0 M_3^2 + 0.6 m_{\tilde{t}}^2   + 0.2 A_t M_3 \label{eq:mstop} \\
M_3 (M_Z) & \simeq & 3 M_3 \label{eq:M3} \\
A_t (M_Z) & \simeq & - 2.3 M_3 + 0.2 A_t. \label{eq:At}
\eea 
In the case of $m_{\tilde{t}}$ the coefficients represent averages of  exact coefficients that would 
appear in separate expressions for $ m_{\tilde{t}_L}^2$ and $m_{\tilde{t}_R}^2$.

In the limit when the stop mass, $m_{\tilde{t}} (M_Z) \simeq 300$ GeV, originates mainly from $M_3$, from Eq.~(\ref{eq:mstop}) we see we need $M_3 \simeq 130$ GeV.
Then Eq.~({\ref{eq:At}) shows that the necessary $|A_t(M_Z)| \simeq 500$ GeV 
is obtained only when $A_t \lsim -1000$ GeV or $A_t \gsim 4000$ GeV at the GUT scale, 
in both cases it has to be signifficantly larger than other SSBs.
The contribution from the terms in Eq. (\ref{eq:MZ_gut}) containing $M_3$ and $A_t$ is at least $(600 \, {\rm GeV})^2$ and therefore large radiative correction has to be 
cancelled either by $\mu^2$ or $m_{H_u}^2 (M_{GUT})$. 
If $m_{\tilde{t}}$ is not negligible at the GUT scale, $M_3$ can be 
smaller, but in this case we need even larger $A_t$ and the conclusion is basically the same. 
The situation improved by considering large $A_t$ term. However, we still need at least 3 $\%$ fine tuning.

Although $M_Z$ results from cancellations between SSBs~\footnote{To some extend this is already 
signaled by bounds on masses of superpartners from direct searches, however the limits on the Higgs mass 
and the above discussion make this absolutely clear. }
it does not mean that it is necessarily fine tuned. 
SUSY breaking scenarios typically  produce SSBs 
which are related to each other in a specific way 
in which case 
we should not treat each one of them separately.
Although, in this case, our conclusions about the level of fine tuning 
are irrelevant, 
the discussion above tells us 
what relations between SSBs have to be generated, 
should the $M_Z$ emerge in a natural way.
For instance, it was recently discussed that fine tuning can be reduced with a proper mixture of  anomaly and modulus mediation
\cite{Choi:2005hd,Kitano:2005wc,Lebedev:2005ge} which produces boundary conditions leading to large stop mixing at the EW scale and an initial value of $m_{H_u}^2$ canceling most of the contribution from running.

Even if SUSY breaking scenario produces SSBs related to each other in a way that guarantees large degree of cancellation, still  they cannot be arbitrarily heavy because in that case the $M_Z$ much smaller than superpartner masses would emerge as a coincidence and we would not have a natural explanation for it. This ``coincidence" problem is further amplified by the fact that the relations that have to be satisfied between SSBs in order to recover the correct $M_Z$ depend on the energy interval they are going to be evolved. Therefore a SUSY breaking scenario would have to  know that SSBs will evolve according to MSSM RG equations, and exactly from $M_{GUT}$ to $M_Z$. 

There is one possibility which to large extend overcomes this problem.
If we allow negative stop masses squared at the GUT scale several interesting things happen simultaneously. First of all, from Eq.~(\ref{eq:mstop}) we see that unless $m_{\tilde{t}}$ is too large compared to $M_3$ it will run to positive values at the EW scale. At the same time the contribution to $m_{H_u}^2$ from the energy interval where $m_{\tilde{t}}^2 <0 $ partially or even exactly cancels the contribution from the energy interval where $m_{\tilde{t}}^2 >0 $ and so the EW scale value of $m_{H_u}^2$ can be arbitrarily close to the starting value at $M_{GUT}$, see Fig.~\ref{fig:RGrunning}. 
From Eq.~(\ref{eq:MZ_gut}) we see that this happens for $m_{\tilde{t}}^2 \simeq - 4 M_3^2$ (neglecting $A_t$).
No cancellation between initial value of $m_{H_u}^2$ (or $\mu$) and the contribution from the running is required.
And finally, from Eqs.~(\ref{eq:mstop}) and  (\ref{eq:At}) we see that the stop mixing is typically much larger than in the case with positive stop masses squared. For positive (negative) stop masses squared we find  $|A_t(M_Z) / m_{\tilde{t}} (M_Z)| \lsim 1$  $(\gsim 1)$
starting with $A_t = 0$ and small $ m_{\tilde{t}}$ at the GUT scale. Starting with larger $ m_{\tilde{t}}$ the mixing is even smaller (larger) in the positive (negative) case. Therefore large stop mixing at the EW scale is generic in this scenario and actually it would require very large GUT scale values of $A_t $ to end up with small mixing at the EW scale.
\begin{figure}
\includegraphics[width=3.in]{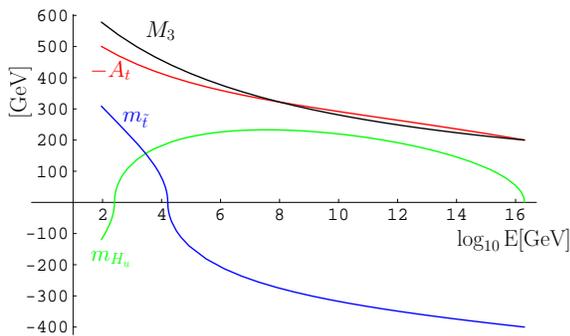}
\caption{Renormalization group running of relevant SSBs for $\tan \beta = 10$ and GUT scale boundary conditions: $-A_t = M_3 = 200$ GeV, $m_{\tilde{t}}^2 = -(400 \, {\rm GeV})^2$ and $m_{H_u}^2 = 0 \, {\rm GeV}^2$.
In order to have both mass dimension one and two parameters on the same plot and keep information about signs, we define $m_{H_u} \equiv m_{H_u}^2/\sqrt{|m_{H_u}^2|} $ and $m_{\tilde{t}}\equiv m_{\tilde{t}}^2/\sqrt{|m_{\tilde{t}}^2|} $. }
\label{fig:RGrunning}
\end{figure}

It turns out that in the region where $m_{H_u}^2$ gets negligible contribution from running, the radiatively generated stop mixing is close to maximal even when starting with negligible mixing at the GUT scale. In this case, comparing Eqs.~(\ref{eq:mstop}) and (\ref{eq:At}), we find~\footnote{To be more precise the generated mixing is somewhat larger than that shown in this equation, since we should minimize the potential at the SUSY scale $\sim m_{\tilde{t}}$ and should not run SSBs all the way to $M_Z$ (see~Fig.~\ref{fig:RGrunning}).}
\begin{equation}
A_t(M_Z) / m_{\tilde{t}} (M_Z) \simeq -1.5 + 0.2 A_t/M_3.
\label{eq:generatedAt}
\end{equation}
Slightly more negative stop masses squared at the GUT scale would result in maximal stop mixing at the EW scale even when starting with negligible $A_t$. Nevertheless
the example in~Fig.~\ref{fig:RGrunning} with simple GUT scale boundary conditions  already
leads to EW scale parameters
$m_{\tilde{t}}(M_Z) \simeq 300$ GeV and $A_t(M_Z) = -500$ GeV producing sufficiently heavy  Higgs boson, $m_h \simeq 115.4$ GeV. Small variations of GUT scale parameters, including positive or negative values of $m_{H_u}^2$,  would produce similar results and scaling all parameters up would lead to larger Higgs mass.
 
In a theory which predicts $m_{\tilde{t}}^2 \simeq - 4 M_3^2$, the
fine tuning problem is entirely solved. The contribution to $m_{H_u}^2$ from the running is negligible and the ${\cal O}(M_Z^2)$ values of $m_{H_u}^2$ and $\mu^2$ at the GUT scale naturally result in the correct $M_Z$. However, the absence of fine tuning is quite robust and the relation above does not have to be satisfied very precisely.  If we define $\alpha$ by
\bea \f{|m_{\tilde{t}}|}{M_3} & = &
2(1+\alpha ), \eea
then the EW scale (\ref{eq:MZ_gut}) can be written as 
\bea M_Z^2 & \simeq & -
1.9\mu^2 - 1.2 m_{H_u}^2  -12 \alpha M_3^2. 
\eea
We see that requiring 
fine tuning less than $10 \%$, large range of $\alpha$
is allowed (for $M_3 \simeq 200$ GeV):
\bea -0.17 < \alpha < 0.17. \eea
This interval is shrinking with increasing $M_3$ which is a sign of the coincidence problem discussed above.

In summary, a 
very reasonable set of SSBs at the GUT scale: 
$M_3 \gsim 200$ GeV, $|m_{\tilde{t}_L}| \simeq |m_{\tilde{t}_R}| \simeq (1.7 \, - \, 2.3) M_3$ and $A_t$ of order the other SSBs or smaller naturally reproduces the correct EW scale. The EW scale value of $m_{H_u}^2$ is very close to the starting value at the GUT scale.  In a simplified way this can be understood as effectively lowering the scale where SSBs are generated to the scale where $m_{\tilde{t}} \simeq 0$ (in the example in Fig.~\ref{fig:RGrunning} it is $10$ TeV).
From this scale SSBs run in a similar way they would run when starting with positive stop masses. However this scale is much closer to the EW scale and so $\delta m_{H_u}^2$, Eq.~(\ref{eq:del_mhu}), generated between this scale and the EW scale is considerably smaller.  
The stop mixing at the EW scale is close to maximal, but it is generated radiatively
starting from a small mixing  at the GUT scale. It is to be compared 
with the positive case which requires $A_t$
to be several times larger than other SSBs in order to produce 
large enough mixing to satisfy LEP bounds on the Higgs mass.
Thus considering  negative values for stop masses squared keeps the desirable 
feature of radiative electroweak symmetry breaking and minimizes 
fine tuning. The Higgs mass is automatically enhanced 
and staying above the LEP bound does not require additional 
constraints on the rest of SUSY parameters. 

However strong constraints can originate when considering possible CCB minima.
At the EW scale all scalar masses squared (except $m_{H_u}^2$) are positive, nevertheless, as already discussed,
very large $A_t$ term would generate
a CCB minimum at around the EW scale~\cite{Gunion:1987qv,Casas:1995pd}.
Then the EW vacuum should be the global minimum since otherwise
the EW vacuum would rapidly tunnel to the CCB minimum as
the barrier is neither high nor thick. 
The optimal sufficient condition to avoid a
CCB vacuum in ($H_u, \tilde{t}_L, \tilde{t}_R$) plane
is $|A_t| \lsim 2 m_{\tilde{t}}$~ \cite{LeMouel:2001ym}.
The generated $A_t$ in the region we consider~(\ref{eq:generatedAt}) may be close but is typically well within this bound.  

Negative stop masses squared at the GUT scale result in unbounded
from below (UFB) potential along the D-flat direction
\cite{Frere:1983ag,Derendinger:1983bz}.
The tree level
potential at the GUT scale gets large loop corrections and the RG
improved effective potential is no longer UFB. 
However, it generates a large VEV (compared to the EW scale)
CCB minimum. If the potential energy of the CCB minimum is lower
than that of the EW minimum, the EW minimum can tunnel to the CCB minimum.
In most of parameter space the tunneling rate is too small and the EW
vacuum can live longer than the age of the universe
\cite{Riotto:1995am,Kusenko:1996jn}.
More precisely, the longetivity of the
metastable EW vacuum puts a constraint 
$m_{\tilde{t}}(M_Z) \gsim \f{1}{10} M_3(M_Z)$ \cite{Riotto:1995am}, 
and again the region of parameter space we consider is entirely safe from this bound (nevertheless it tells us that stop masses squared cannot be arbitrarily large and negative at the GUT scale).

A possible problem is that after inflation
the universe is likely
to settle down in a large VEV CCB vacuum
rather than the EW vacuum. This is worrisome 
since the tunneling rate to the EW vacuum would be very small.
However, if the reheating temperature is high enough, the large VEV CCB minimum
might disappear in finite temperature effective potential.
For a given set of SSBs, 
there is a minimum reheating temperature
above which the large VEV CCB vacuum disappears~\cite{Falk:1996zt}.
It depends on how inflation ends
and SSBs will constrain compatible
inflation scenarios.


In this letter we focused on the SUSY parameters relevant 
for radiative EWSB and discussion of fine tuning. 
An interesting signature of this scenario is stop splitting, 
$m_{\tilde{t}_1,\tilde{t}_2} \simeq m_{\tilde{t}} (M_Z) \mp m_{t}$, 
and stops considerably lighter than gluino:
$m_{\tilde{t}}(M_Z) \lsim 0.5 m_{\tilde g}$ with $m_{\tilde g} \gsim 600$ GeV (the lighter stop thus can be as light as 130 GeV).  
Besides these the scenario has a light Higgsino, a possible candidate for LSP.

Other scalar soft masses squared are 
unconstrained by considerations of fine tuning and 
can be positive or even all negative at the GUT scale 
in complete models. In some SUSY breaking scenarios the sign of the scalar masses squared is not determined~\cite{Feng:2005xb} while in others it can be fixed and negative. For example,
negative scalar masses squared arise in gauge mediation with gauge fields as messengers~\cite{Giudice:1997ni} or one can utilize the minus sign arising in the see-saw mechanism for scalar masses~\cite{Kim:2005qb}.
It is desirable to build fully realistic 
models of this type 
in which constrains from CCB can be addressed. In specific scenarios
additional potential problems may occur, 
like negative slepton masses squared at the EW scale 
since the right-handed sleptons  receive contribution only from $M_1$. 
Even if this contribution is large enough to drive sleptons positive, 
we still can end up with stau LSP.  

Nevertheless all the positive features of negative stop masses squared 
suggest it is worthwhile to seriously search for  models which 
can lead to boundary conditions discussed above.


\vspace{0.2cm}
We thank K. Agashe, A. Falkowski, G. Giudice and R. Rattazzi for discussions.
HK thanks CERN and
IAS for the hospitality during the visit. This work was supported
by the U.S. Department of Energy, grant DE-FG02-90ER40542, by the ABRL Grant No. R14-2003-012-01001-0, the BK21 of
MOE, Korea and the SRC of the KOSEF through
the CQUeST of Sogang University
with grant number R11-2005-021.



\end{document}